\documentclass[%
 reprint,
superscriptaddress,
longbibliography,
 amsmath,amssymb,
 aps,
]{revtex4-1}

\usepackage[normalem]{ulem}
\usepackage{color}
\usepackage{graphicx}
\usepackage[raggedright]{subfigure}
\usepackage{dcolumn}
\usepackage{bm}

\newcommand{\bologna}{\affiliation{Department of Physics and Astrophysics, University of Bologna and INFN-Bologna, 40126 Bologna, Italy}}

\newcommand{\chicago}{\affiliation{Department of Physics \& Kavli Institute of Cosmological Physics, University of Chicago, Chicago, IL 60637, USA}}

\newcommand{\coimbra}{\affiliation{LIBPhys, Department of Physics, University of Coimbra, 3004-516 Coimbra, Portugal}}

\newcommand{\columbia}{\affiliation{Physics Department, Columbia University, New York, NY 10027, USA}}

\newcommand{\lngs}{\affiliation{INFN-Laboratori Nazionali del Gran Sasso and Gran Sasso Science Institute, 67100 L'Aquila, Italy}}

\newcommand{\mainz}{\affiliation{Institut f\"ur Physik \& Exzellenzcluster PRISMA, Johannes Gutenberg-Universit\"at Mainz, 55099 Mainz, Germany}}

\newcommand{\heidelberg}{\affiliation{Max-Planck-Institut f\"ur Kernphysik, 69117 Heidelberg, Germany}}

\newcommand{\munster}{\affiliation{Institut f\"ur Kernphysik, Westf\"alische Wilhelms-Universit\"at M\"unster, 48149 M\"unster, Germany}}

\newcommand{\nikhef}{\affiliation{Nikhef and the University of Amsterdam, Science Park, 1098XG Amsterdam, Netherlands}}

\newcommand{\nyuad}{\affiliation{New York University Abu Dhabi, Abu Dhabi, United Arab Emirates}}

\newcommand{\purdue}{\affiliation{Department of Physics and Astronomy, Purdue University, West Lafayette, IN 47907, USA}}

\newcommand{\rpi}{\affiliation{Department of Physics, Applied Physics and Astronomy, Rensselaer Polytechnic Institute, Troy, NY 12180, USA}}

\newcommand{\rice}{\affiliation{Department of Physics and Astronomy, Rice University, Houston, TX 77005, USA}}

\newcommand{\stockholm}{\affiliation{Oskar Klein Centre, Department of Physics, Stockholm University, AlbaNova, Stockholm SE-10691, Sweden}}

\newcommand{\subatech}{\affiliation{SUBATECH, IMT Atlantique, CNRS/IN2P3, Universit\'e de Nantes, Nantes 44307, France}}

\newcommand{\torino}{\affiliation{INFN-Torino and Osservatorio Astrofisico di Torino, 10125 Torino, Italy}}

\newcommand{\ucla}{\affiliation{Physics \& Astronomy Department, University of California, Los Angeles, CA 90095, USA}}

\newcommand{\ucsd}{\affiliation{Department of Physics, University of California, San Diego, CA 92093, USA}}

\newcommand{\wis}{\affiliation{Department of Particle Physics and Astrophysics, Weizmann Institute of Science, Rehovot 7610001, Israel}}

\newcommand{\zurich}{\affiliation{Physik-Institut, University of Zurich, 8057  Zurich, Switzerland}}

\newcommand{\paris}{\affiliation{LPNHE, Université Pierre et Marie Curie, Université Paris Diderot, CNRS/IN2P3, Paris 75252, France}}

\newcommand{\freiburg}{\affiliation{Physikalisches Institut, Universität Freiburg, 79104 Freiburg, Germany}}

\begin{document}

\title{Search for WIMP Inelastic Scattering Off Xenon Nuclei With XENON100}
\author{E.~Aprile}\columbia

\author{J.~Aalbers}\nikhef

\author{F.~Agostini}\lngs\bologna

\author{M.~Alfonsi}\mainz

\author{F.~D.~Amaro}\coimbra

\author{M.~Anthony}\columbia

\author{F.~Arneodo}\nyuad

\author{P.~Barrow}\zurich

\author{L.~Baudis}\email[E-mail: ]{laura.baudis@uzh.ch}\zurich

\author{B.~Bauermeister}\stockholm

% ...

\collaboration{XENON Collaboration}\email[E-mail: ]{xenon@lngs.infn.it}\noaffiliation

\date{\today}

\begin{abstract}
We present the first constraints on the spin-dependent, inelastic scattering cross section of Weakly Interacting Massive Particles (WIMPs) on nucleons from XENON100 data with an exposure of 7.64$\times$10$^3$\,kg\,day. XENON100 is a dual-phase xenon time projection chamber with 62\,kg of active mass, operated at the Laboratori Nazionali del Gran Sasso (LNGS) in Italy and designed to search for nuclear recoils from WIMP-nucleus interactions. Here we explore inelastic scattering, where a transition to a low-lying excited nuclear state of $^{129}$Xe is induced. The experimental signature is a nuclear recoil observed together with the prompt de-excitation photon. We see no evidence for such inelastic WIMP-$^{129}$Xe interactions. A profile likelihood analysis allows us to set a 90\% C.L. upper limit on the inelastic, spin-dependent WIMP-nucleon cross section of $3.3 \times 10^{-38}$\,cm$^{2}$  at 100\,GeV/c$^2$.  This is the most constraining result to date, and sets the pathway for an analysis of this interaction channel in upcoming, larger dual-phase xenon detectors.

\end{abstract}

\maketitle


\section{\label{sec:intro} Introduction}

Astrophysical and cosmological evidence indicates that the dominant mass fraction of our Universe consists of some yet unknown
form of dark, or invisible matter. The dark matter could be made of stable or long-lived and yet undiscovered particles. Well-motivated
theoretical models beyond the Standard Model of particle physics predict the existence of Weakly Interacting Massive
Particles (WIMPs), which are natural candidates for dark matter. This hypothesis is currently being tested by several direct
and indirect detection experiments, as well as at the LHC~\cite{Bertone:2010zza,Baudis:2016qwx}.

Most direct detection searches focus on elastic scattering of galactic dark matter particles off nuclei, where the keV-scale 
nuclear recoil energy is to be detected~\cite{Undagoitia:2015gya,Baudis:2015mpa}. In this work, the 
alternative process of inelastic scattering is explored, where a WIMP-nucleus scattering induces a transition to a low-lying 
excited nuclear state. The experimental signature is a nuclear recoil detected together with the prompt de-excitation 
photon~\cite{Ellis:1988nb}. 

We consider the $^{129}\text{Xe}$ isotope, which has an abundance of 26.4\% in natural xenon, and a lowest-lying 
3/2$^{+}$ state at 36.9\,keV above the 1/2$^+$ ground state. The electromagnetic nuclear decay has a half-life of 0.97\,ns. 
The signatures and structure functions for inelastic scattering in xenon have been studied in detail in~\cite{Baudis:2013bba}. It was found that this channel is complementary to spin-dependent, elastic scattering, and that it dominates the integrated rates above $\simeq10$\,keV of deposited energy. 
In addition, in case of a positive signal, the observation of inelastic scattering would provide a clear 
indication of the spin-dependent nature of the fundamental interaction. 

This paper is structured as follows.  In Section~\ref{sec:xenon100} we briefly describe the main features of the XENON100 detector.  In Section~\ref{sec:analysis} we introduce the data 
sets employed in this analysis and detail the data analysis method, including the simulation of the expected signal and the 
background model. We conclude in Section~\ref{sec:results} with our results, and discuss the new constraints on inelastic WIMP-nucleus interactions.

\section{The XENON100 Detector}
\label{sec:xenon100}

The XENON100 experiment operates a dual-phase (liquid and gas) xenon time projection chamber (TPC) at the Laboratori Nazionali 
del Gran Sasso (LNGS) in Italy. It contains 161\,kg of xenon in total,  with  62\,kg in the active region of the TPC. These 
are monitored by 178 1-inch square, low-radioactivity, UV-sensitive photo-multiplier tubes (PMTs) arranged in two arrays, one in the liquid 
and one in the gas. The PMTs detect the prompt scintillation (S1) and the delayed, proportional scintillation signal (S2) 
created by a particle interacting in the active TPC region. The S2-signal is generated due to ionization electrons, drifted 
in an electric field of 530\,V/cm and extracted into the gas phase by a stronger field of $\sim$ 12\,kV/cm, where the proportional scintillation, or electroluminiscence, 
takes place.
The horizontal position, $(x,y)$, of the interaction site is reconstructed from the position of the S2 shower, while the depth of the interaction, $z$, is given by the drift time measurement.
The TPC thus yields a three-dimensional event localization, with an $(x,y)$ resolution of $<$3\,mm (1\,$\sigma$), and a $z$ resolution of  $<$0.3\,mm (1\,$\sigma$), enabling to reject the majority of background events via fiducial volume selections~\cite{Aprile:2011dd}. The ratio S2/S1 provides the basis for distinguishing between nuclear recoils (NRs), as induced by fast neutrons and expected from elastic WIMP-nucleus scatters, and electronic recoils (ERs) produced by $\beta$- and $\gamma$-rays.  A 4\,cm thick liquid xenon (LXe) layer surrounds the TPC and is monitored by 64 1-inch square PMTs, providing an effective active veto for further background reduction.

XENON100 has acquired science data between 2009-2015, and has set competitive constraints on spin-independent~\cite{Aprile:2012nq,Aprile:2016swn} 
and spin-dependent~\cite{Aprile:2013doa,Aprile:2016swn} elastic WIMP-nucleus 
scatters, on solar axions and galactic ALPs~\cite{Aprile:2014eoa}, as well as on leptophilic dark matter models~\cite{Aprile:2015ade,Aprile:2015ibr,Aprile:2017yea}. 

Here we explore a new potential dark matter interaction channel in the XENON100 detector, caused by spin-dependent, inelastic WIMP-$^{129}$Xe interactions. The expected inelastic scattering signature is a combination between an ER and a NR, due to the short lifetime of the excited nuclear state and  the short mean free path of $\sim$0.15\,mm of the 39.6\,keV de-excitation photon.

\begin{figure*}[t!]
	\subfigure[]{\includegraphics[width=0.49\linewidth]{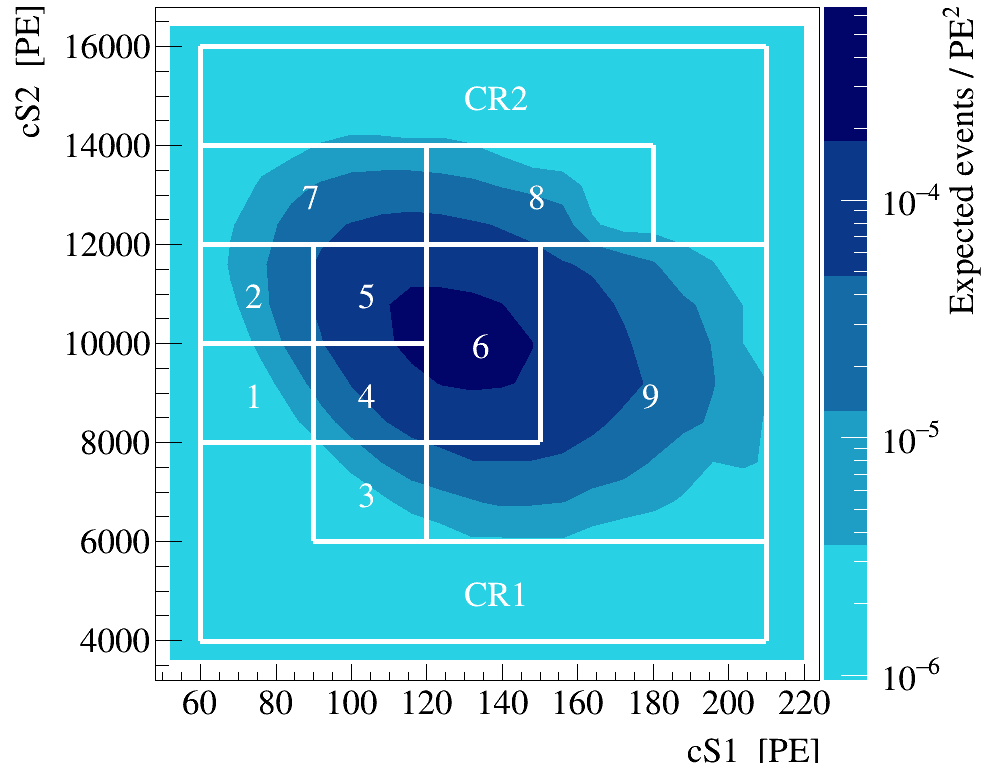}}
	\subfigure[]{\includegraphics[width=0.49\linewidth]{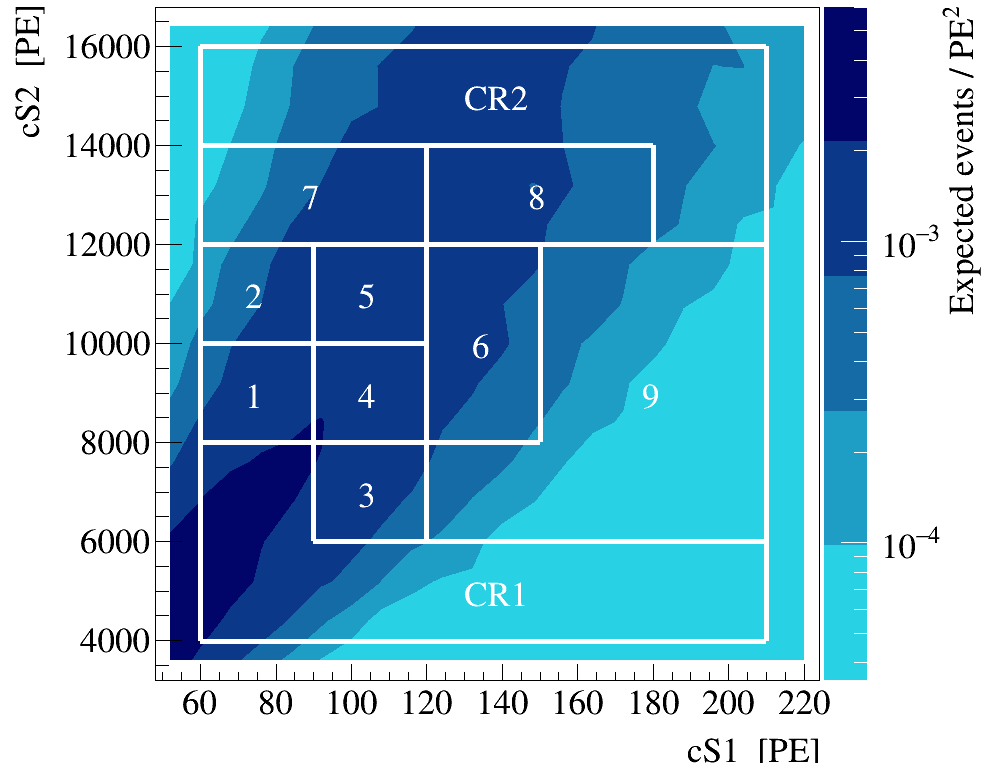}}
	\caption{Signal (1-9) and control (CR1 and CR2) regions for the inelastic WIMP-$^{129}$Xe search in the (cS2,cS1)-plane.
		Figure (a) shows the signal distribution for a simulated WIMP of mass 100\,GeV/c$^2$ normalized to 50\,events, while Figure (b) is obtained using  
		normalized $^{60}$Co calibration data and represents the background expectation distribution.}
  \label{fig:SR}
\end{figure*}

\section{Data Analysis}
\label{sec:analysis}

This analysis is performed using XENON100 Run-II science data, with 224.6~live days of data taking. The detector's response to ERs has been characterized with $^{60}$Co and $^{232}$Th calibration sources, while the response to NRs was calibrated with an $^{241}$AmBe $(\alpha,n)$-source. The fast neutron from the latter gives rise to elastic and inelastic neutron-nucleus scatters, and can thus be employed to define the expected signal region for inelastic WIMP-nucleus scatters.

\subsection{Signal Correction} 

A particle interaction in the liquid xenon produces an S1 and a correlated S2 signal with a certain number of photoelectrons (PE) observed by the PMTs. The non-uniform scintillation light collection by the PMT arrays, due to solid angle effects, Rayleigh scattering length, reflectivity, transmission of the electrodes, etc, lead to a position-dependent S1 signal. The warping of the top meshes (inducing a variation in the width of the gas gap between the anode and the liquid-gas interface), the absorption of electrons by residual impurities as they drift towards the gas region, as well as solid angle effects, lead to a position-dependent S2 signal. These signals are thus corrected in 3 dimensions, using various calibration data, as detailed in~\cite{Aprile:2011dd,Aprile:2012vw}, with the corrected quantities denoted as cS1 and cS2, defined in~\cite{Aprile:2012vw}. 

\subsection{Signal Region and Event Selection} 

The inelastic scattering of a WIMP with a $^{129}$Xe nucleus is expected to produce an energy deposit via a NR with the subsequent emission of  
a 39.6\,keV de-excitation photon. The largest fraction of the energy released in the event is via the ER, due to the emitted photon which loses its energy in the LXe.
This represents an unusual signature compared to the one expected from an elastic scatter, and makes the signal region to overlap the ER background region.
The region of interest (ROI) selected for this analysis surrounds the 39.6\,keV xenon line in the (cS1,cS2)-plane 
and is based on $^{241}$AmBe calibration data, where such inelastic scatters are induced by fast neutrons. The ROI extends from 60 to 210\,PE in cS1, from 4$\times \, 10^3$ to 16$\times \, 10^3$\,PE in cS2 and is further divided  into sub-regions as shown in Figure~\ref{fig:SR}.  These sub-regions were defined to  contain a (roughly) similar number of expected background events in each region. The control regions (denoted as CR1 and CR2 in the figures), are selected to be as close as possible to the ROI,  and are used for cross checks of the background shape distribution.

Apart from the condition to occur in the defined ROI, valid events are required to fulfil several selection criteria, 
which can be summarized as follows: basic data quality cuts, energy selection and S2 threshold cut, veto cut for events with energy release in the detector's 
active LXe shield, selection of single-scatter events and of a predefined fiducial volume of 34\,kg.  
Our analysis closely follows the event selection criteria described in detail in \cite{Aprile:2012vw} for Run-II, with the following few exceptions. 
The cut on the width of the S2 signal as a function of drift time (where the maximal drift time is 176\,$\mu$s and the width values range from $\sim$1-2\,$\mu$s) has been optimized on a sample of events selected from the 39.6\,keV line and set to a 95\% acceptance of these. This cut ensures that the broadening of S2-signals due to diffusion is consistent with the $z$-position calculated from the observed time difference between the S1 and S2 signals. Events are required to be single-scatters by applying a threshold cut on the size of the 
second largest S2 peak. For this analysis, the threshold has been optimized to 160\,PE and is constant with respect to S2 signal size.

\subsection{Signal Simulation} 
\label{sec:signal}

\begin{figure*}[t!]
	\subfigure[]{\includegraphics[width=0.49\linewidth]{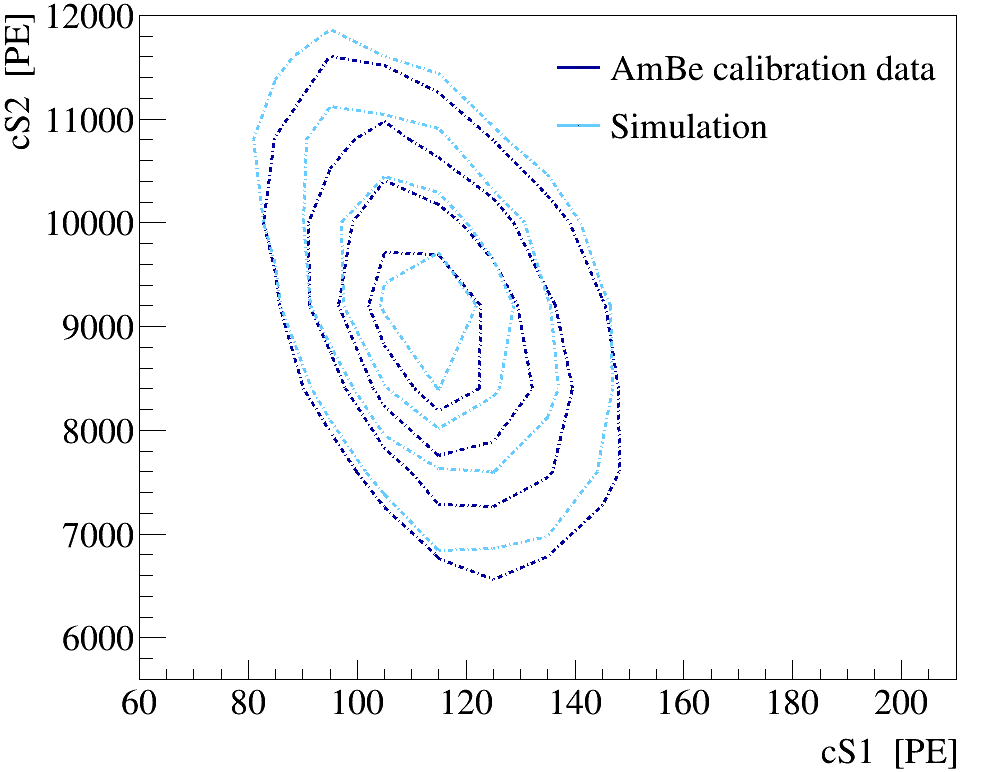}}
	\subfigure[]{\includegraphics[width=0.49\linewidth]{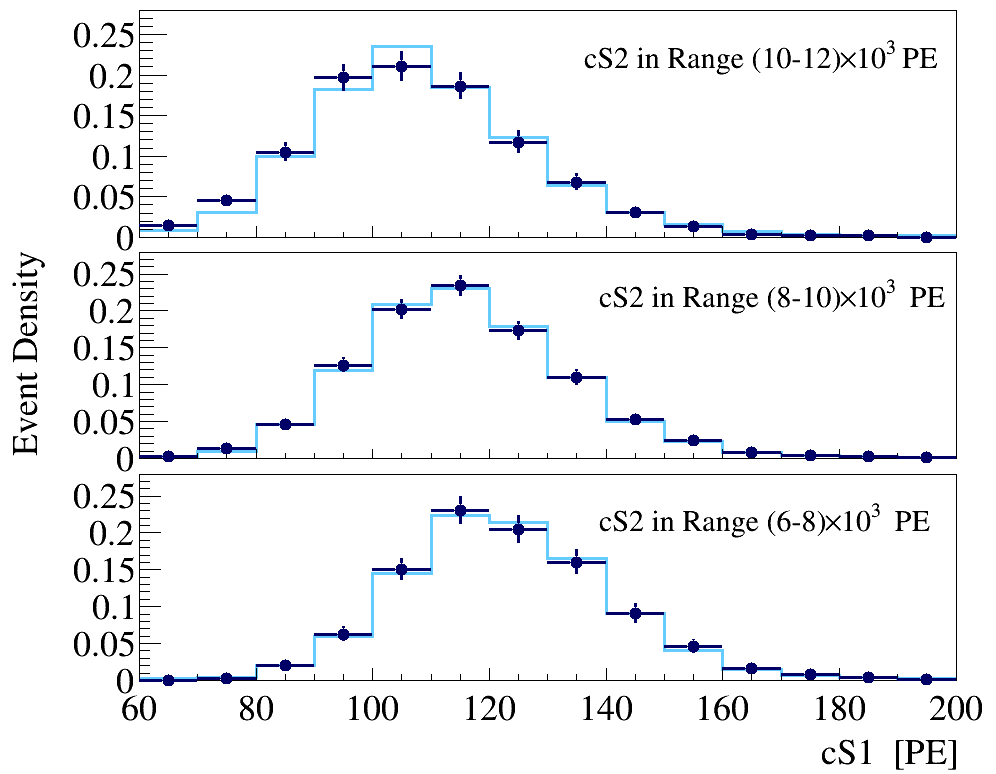}} 
	\caption{A simulation of the 39.6\,keV  xenon line activated from $^{241}$AmBe neutrons is compared with  data.
		 Figure~(a) compares contours of equal density in the (cS1,cS2)-plane, while Figure~(b) shows the same distribution projected in cS1 for
		 several ranges of cS2, the histograms are normalized to unit area. Light and dark blue represent simulation and data, respectively.
		}
		
  \label{fig:mc_comp}
\end{figure*}

The detector response to inelastic WIMP-$^{129}$Xe interactions was simulated using an empirical signal model, described in this section.
The total deposited energy is divided into two independent contributions: one coming from the 39.6\,keV de-excitation photon and the other  from  
the simultaneous nuclear recoil of the xenon atom. The detected light (S1) and charge (S2) signals are simulated separately for each of the two contributions 
and then added together. This recipe has been followed  because the light and charge yields depend both on the type of interaction (ER vs. NR), and on the deposited energy.

The distribution of an ER induced by the de-excitation photon in the (cS1,cS2)-plane  is simulated assuming a two dimensional normal probability distribution function (pdf), $f(\rm{cS1}_{er},\rm{cS2}_{er})$, 
described (apart from a constant normalization factor) by the following equation:

\begin{multline}
	f(cS1_{er},cS2_{er})  = {\rm exp} \Big\{ -\frac{1}{2(1-\rho^2)} \Big[ \frac{(cS1_{er} - \mu_{cS1})^2}{\sigma_{cs1}^2} + \\ 
	 \frac{(cS2_{er} - \mu_{cS2})^2}{\sigma_{cs2}^2} - \frac{2\rho\cdot(cS1_{er} - \mu_{cs1}) (cS2_{er} - \mu_{cs2})} {\sigma_{cs1}\sigma_{cs2}} \Big] \Big\}
\label{f:2dgaus}
\end{multline}
where $\mu_{cS1}$ and $\mu_{cS2}$ 
represent the average observed $\rm{cS1_{er}}$ and $\rm{cS2_{er}}$ signals given a 39.6\,keV ER, $\sigma_{cs1}$ and $\sigma_{cs2}$ are the standard deviation in $\rm{cS1_{er}}$ and $\rm{cS2_{er}}$ respectively,
while $\rho$ stands for the correlation between the cS1 and cS2 signals. The detector-related light yield $L_y$  at 39.6\,keV, necessary to evaluate the average number of prompt photons detected 
($\mu_{cS1}$), is obtained from the NEST model~\cite{NEST,Geant1,Geant2} fit to data collected with several $\gamma$-lines.
The same model is used to predict the charge yield at 39.6\,keV, which is then scaled according to the detector's secondary scintillation gain $Y$. 
 The latter is determined from the detector's response to single electrons~\cite{SingleE}.
The energy resolution at 39.6\,keV in cS1 and cS2 has been measured to be 15.8\% and 14.7\%, respectively, and is used to extract the standard 
deviations $\sigma_{cs1}$, $\sigma_{cs2}$.  The correlation parameter is measured
using the 164\,keV line from the decay of the $^{131m}$Xe isomer ($T_{1/2}$=11.8\,d) produced during the  $^{241}$AmBe calibration. This $\gamma$-line is chosen since, unlike the 39.6\,keV line, 
it is not associated with a NR and a measure of the (cS1,cS2) correlation of a pure ER interaction is possible. The correlation coefficient however depends on energy due to 
electron recombination effects. Its measured value at 164\,keV is thus corrected based on the NEST expected recombination fractions for those energies. 
The corrected correlation coefficient is then $\rho \, = \, -0.4 \pm 0.1$.

The cS1 and cS2 distributions from the NR contribution are predicted starting from the expected nuclear recoil energy spectrum
of WIMP inelastic interactions~\cite{Baudis:2013bba}. The average cS1 and cS2 are given by equations~\ref{f:cs1} and~\ref{f:cs2} respectively:
\begin{equation}
cS1_{nr} ~=~ E_{nr} \, \mathcal{L}_{\text{eff}}(E_{nr}) \, L_{y} \, \frac{S_{nr}}{S_{ee}}
\label{f:cs1}
\end{equation}

\begin{equation}
cS2_{nr}  ~ = ~ E_{nr} \, Q_{Y}(E_{nr}) \, Y
\label{f:cs2}
\end{equation}
where $\mathcal{L}_{\text{eff}}$ is the liquid xenon scintillation efficiency for NRs relative to 122\,keVee, while $S_{ee} \, = \, 0.58$  and $S_{nr} \, = \, 0.95$ describe the scintillation 
quenching due to the electric field of ER and NRs, respectively~\cite{ScintQuenching}. The parameterization and uncertainties of $\mathcal{L}_{\text{eff}}$ as a function of nuclear
recoil energy $E_{nr}$ are based on existing direct measurements \cite{run8Result}. The light yield for 122\,keV ERs is taken from the same NEST model fit as described above. For cS2, the parameterization 
of $Q_{Y}(E_{nr})$ is taken from \cite{QY}. Finally, all detector related resolution effects are introduced following the prescriptions described in \cite{Aprile:2012vw}.

The pdf of the ER and NR contributions are then convolved  to obtain the overall pdf of the expected signal.
A 2D (cS1 versus cS2) acceptance map is applied to the signal pdf to reproduce data selection effects. Acceptances are computed separately for each selection 
criteria using the $^{241}$AmBe calibration sample. Acceptances of other selections such as the liquid xenon veto cut, and the single-scatter interaction, represent an exception  for which  
a dedicated computation has been performed. The combined acceptance  of all selection criteria in the region of interest averages to $\sim$$(0.80\pm0.05)$. 
Figure~\ref{fig:SR}\,(a) shows an example of fully simulated signal model for a WIMP mass of 100\,GeV/c$^2$, normalized to 50 events.


\subsection{Signal Validation}

The detector response to inelastic WIMP-$^{129}$Xe interactions was simulated using an empirical signal model. 
The procedure, described in detail in section~\ref{sec:signal}, takes advantage of several 
approximations that have been validated extensively. 
The main aim of the cross check was to reproduce the 39.6\,keV xenon line activated from $^{241}$AmBe neutrons with simulation.
For this purpose the NR energy spectrum expected from inelastic neutron-$^{129}$Xe scattering has been obtained via  Monte Carlo techniques, 
where we take into account the  detector response and the non-uniform spatial distribution. The acceptance of analysis selections to this type of interaction 
have been recomputed. In particular, the acceptance to the double scatter cut differs greatly between neutrons and WIMPs scattering. 
Except for acceptances and NR energy spectrum, the simulation has been performed following the recipe described in the main text. Figure~\ref{fig:mc_comp}
shows a comparisons between simulation (light blue) and calibration data (dark blue), contour lines of equal densities are compared in Figure~(a), 
while Figure~(b) shows the cS1 projected distributions for different ranges in cS2. These considerations thus validate our analysis.


\subsection {Background Model}

The background in the region of interest for inelastic scattering is dominated by ERs and due to the residual radioactivity of detector materials, to $^{85}$Kr present in the liquid xenon, as well as due to $^{222}$Rn decays in the liquid~\cite{Aprile:2011vb}. The background contribution from inelastic scatters of radiogenic or cosmogenic neutrons (producing a 39.6\,keV de-excitation line) is negligible thanks to the very low expected neutron scattering rate in the detector~\cite{Aprile:2013tov}.

The expected background is modelled using data from the $^{60}$Co calibration campaign, which are assumed to represent well the background density distribution 
in the (cS1,cS2)-plane. The calibration sample yields  about $2.2\times10^4$ events in the ROI; these are then scaled to the science data according to a measured scale 
factor $\tau_{\text{bkg}}$. This scale factor, which is merely the ratio between the data and calibration sample yields, is measured in the two control regions shown in Figure~\ref{fig:SR} (labelled CR1 and CR2) separately. The two control regions give compatible results and the computed average is $\tau_{\text{bkg}} \, =  \, 0.034 \pm 0.002 $, where the reported uncertainty 
is of statistical nature only.

The distribution of the calibration sample has been compared to the data of the science run in the two control regions,
and agreement was found within statistical uncertainties. Furthermore, $^{60}$Co calibration data have been compared in the region of interest to  
data from the $^{232}$Th calibration campaign, and systematic uncertainties assessed based on it.


\subsection{Systematic Uncertainties}

\begin{figure}[t!]
  \includegraphics[width=\linewidth]{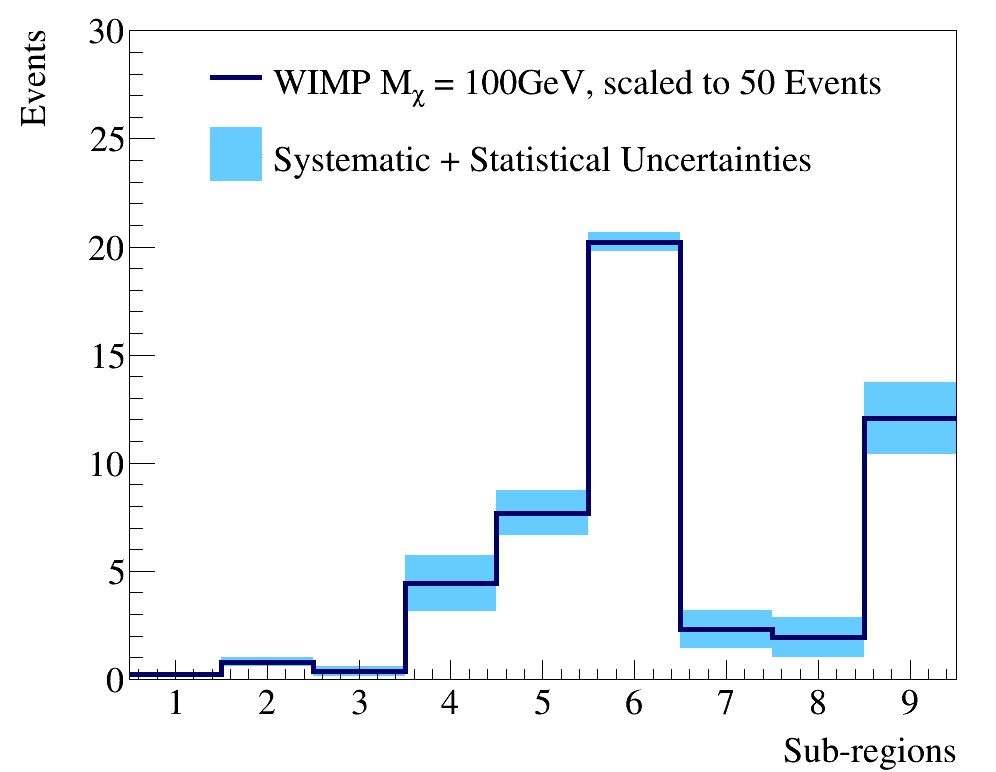}
  \caption{Predicted signal yield in each sub-region (blue curve), along with statistical and systematic uncertainties (shaded area) simulated for a WIMP mass of 100\,GeV/c$^2$.  The signal has been scaled for a total number of 50 events. The sub-regions are defined in Figure~\ref{fig:SR}.}
  \label{fig:unc}
\end{figure}

Uncertainties on the  prediction of the total number of background events arise from the uncertainty on the measurement of the normalisation 
factor, $\tau_{\text{bkg}}$, and amount to 6\%. 
Systematic uncertainty on the shape of the predicted background distribution are assessed by the maximal observed discrepancy in the ROI between
the $^{60}$Co and $^{232}$Th calibration samples. The two sample's normalized yield are compared in each sub-region and the overall largest deviation (incompatible
with statistical fluctuation) is found to be within 4\%.  Consequently, a systematic uncertainty of 4\% is assigned  to the expected yield of each sub-region.
Uncertainties belonging to different sub-regions in the ROI are considered independent from one another.

Uncertainties on the total yield of signal events arising from selections are found to be only very weakly dependent on 
the WIMP mass, and an overall 6\% acceptance uncertainty is applied to all WIMP hypotheses. 

Uncertainties on the energy scale and, more generally, related to detector responses  are parameterised 
using the respective uncertainties on the measures of $L_y$, $\mathcal{L}_{\text{eff}}$, $Y$, $Q_Y$ and $\rho$. The simulation shows 
that these uncertainties mainly affect the pdf of the signal model in the ROI, and very weakly the total signal yield. 
They are taken into account by simulating several signal pseudo-samples for each WIMP mass, where the pseudo-samples are produced 
by varying the model parameters by their $\pm 1$~standard deviation. 
For each sub-region, an overall uncertainty is then computed by adding in quadrature the residual of each pseudo-sample 
with respect to the nominal. Figure~\ref{fig:unc} shows an example of such a systematic uncertainty computation for a WIMP mass of 100\,GeV/c$^2$.

All the uncertainties discussed here are parameterised within a binned profile likelihood function using the ROOSTAT-ROOFIT framework~\cite{roostat,roofit}.
All the parameters related to systematic uncertainties are assumed to be normally distributed.

\begin{figure}[t]
  \includegraphics[width=\linewidth]{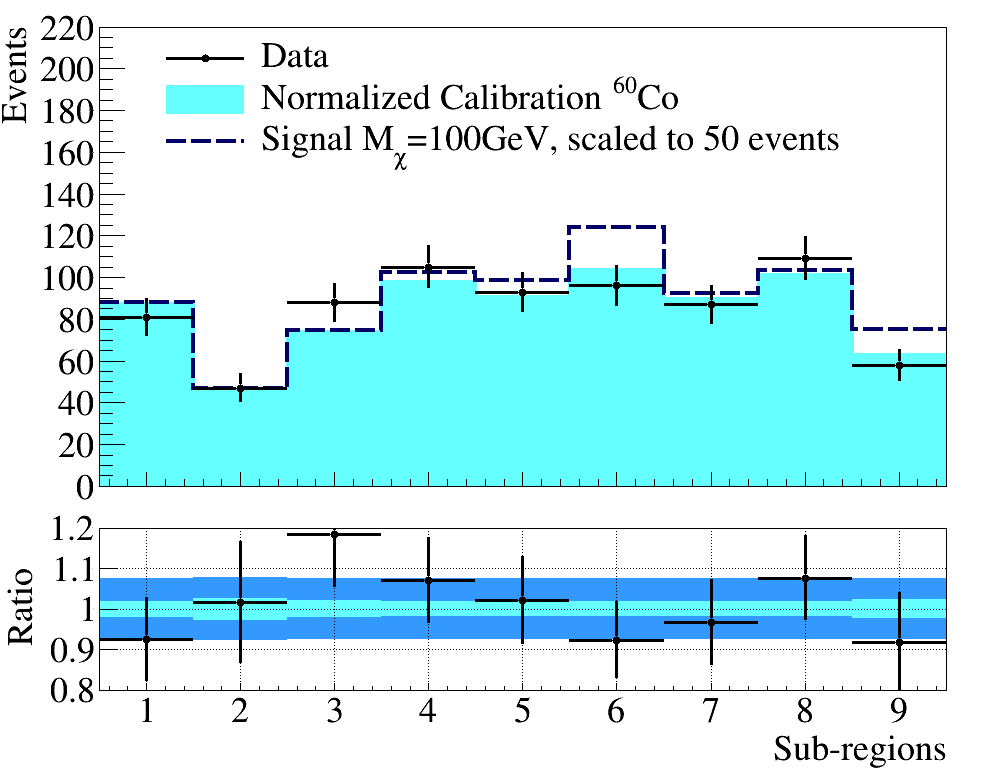}
  \caption{Distribution of  observed events  in the region of interest (data points), along with the normalized distribution from calibration data (filled histogram). The bottom panel displays the ratio
between data and expected background, where the light  and dark blue shaded areas represent the statistical and systematic uncertainty 
on the background expectation, respectively. The expected signal for a WIMP mass of 100\,GeV/c$^2$ (blue dashed), normalized to a total of 50 events, is also shown.}
  \label{fig:dataVSbkg}
\end{figure}

\section{Results and Discussion}
\label{sec:results}

This search is performed using XENON100 Run-II science data, which corresponds to an exposure of 34\,$\times$\,224.6\,kg\,$\cdot$\,days. 
A total of 764 events are observed in the region of interest and no evidence of dark matter can be assessed based on an expected background of
$756 \, \pm \, 5 \,{\rm (stat)} \, \pm 55\, {\rm (syst)}$ events. 
Figure~\ref{fig:dataVSbkg} shows the distribution of  events  in the region of interest, where the bottom panel displays the ratio
between data and expected background. The light and dark blue shaded areas represent the statistical and systematic uncertainty 
on the background expectation, respectively. The expected signal for a WIMP mass of 100\,GeV/c$^2$, normalized to a total of 50 events, is also shown.

\begin{figure}[t]
  \includegraphics[width=\linewidth]{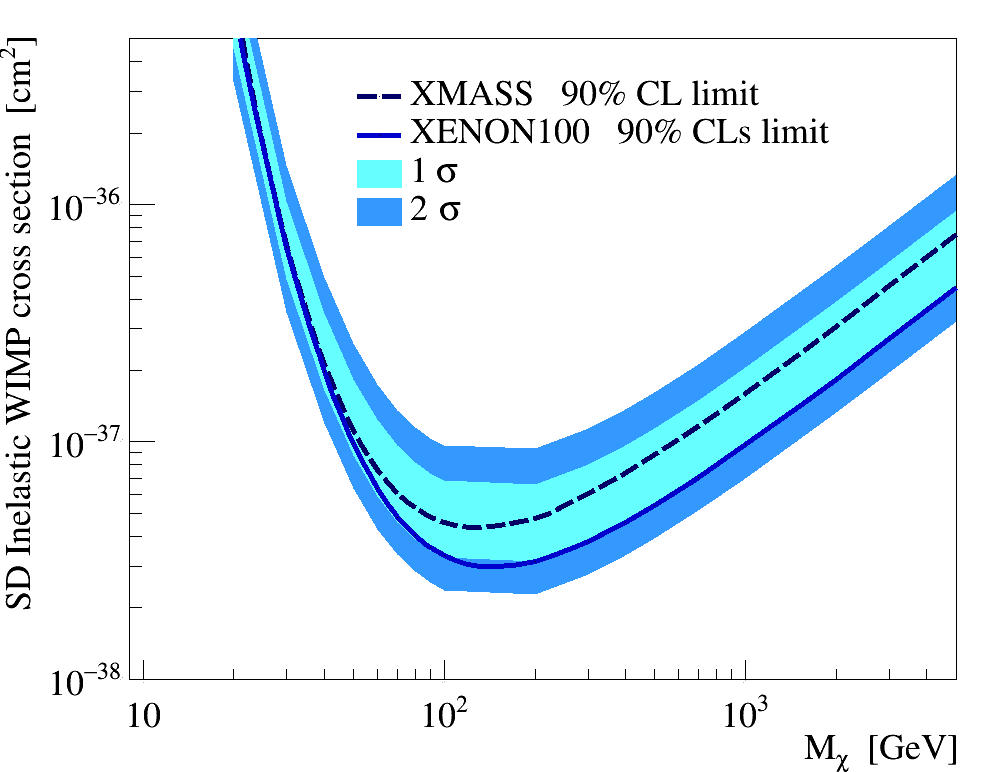}
  \caption{Upper limit (blue curve) on the spin-dependent, inelastic WIMP-nucleon cross section as a function of WIMP mass.  
	  The expected one (light shaded area) and two (dark shaded area) standard deviation uncertainty is also shown. 
	  This result is compared to the upper limit (at 90\% C.L.) obtained by the XMASS experiment (dashed line)~\cite{Uchida:2014cnn}.}
  \label{fig:limits}
\end{figure}

This result is interpreted via a binned profiled likelihood approach by means of the test statistic $\tilde{q}$
and its asymptotic distributions, as  described in \cite{asympt}. 
Assuming  an isothermal WIMP halo with a local density of $\rho_{\chi} \, = \, 0.3$\,GeV/cm$^3$, a local circular velocity of $v_0 \,= \, 220$\,km/s, 
a galactic escape velocity of $v_{\rm{esc}} \, = \, 544$\,km/s~\cite{Smith:2006ym}  
and the nuclear structure factors as computed in~\cite{Baudis:2013bba}, 
a 90\% CL$_s$~\cite{cls} confidence level upper limit is set on the spin-dependent inelastic WIMP-nucleon cross section as a function of the WIMP mass. 

Our result is shown in Figure~\ref{fig:limits}, together with its expected one and two sigma statistical variation.
The most constraining limit is  set for a WIMP of mass 100\,GeV/c$^2$ to a cross section of $3.3 \times 10^{-38}$\,cm$^{2}$ (at 90\% CL$_s$ confidence level). 

This result is compared to the one obtained by the XMASS experiment~\cite{Uchida:2014cnn}, a single phase liquid xenon detector, which used a fiducial volume containing 41\,kg of LXe and 165.9~live days of data.

While these upper limits are not competitive to spin-dependent, elastic scattering results, as obtained by XENON100~\cite{Aprile:2013doa} and LUX~\cite{Akerib:2016lao} 
(bounding the cross section to be $<\,1 \times 10^{-40}$\,cm$^{2}$, at 90\% C.L.,  for a 100\,GeV/c$^2$ WIMP), 
our results are the most stringent for the spin-dependent inelastic channel, and set the stage for a sensitive search of inelastic WIMP-nucleus scattering in running or upcoming liquid xenon experiments such as XENON1T~\cite{Aprile:2015uzo}, XENONnT~\cite{Aprile:2015uzo},  LZ~\cite{Akerib:2015cja}, and DARWIN~\cite{Aalbers:2016jon}. In these larger detectors, with lower intrinsic backgrounds from $^{85}$Kr and $^{222}$Rn decays, and improved self-shielding, the electronic recoil background will be reduced by a few orders of magnitude with respect to XENON100, and ultimately limited by solar neutrino interactions~\cite{Baudis:2013qla}. 
The discovery of this interaction channel would be a clear signature for a spin-dependent nature of the dark matter interaction, and would provide a potential handle  to constrain the WIMP mass with data from one experiment only~\cite{Baudis:2013bba,McCabe:2016aof}.

\section*{Acknowledgments}
We gratefully acknowledge support from the National Science Foundation, Swiss National Science Foundation, Deutsche Forschungsgemeinschaft, Max Planck Gesellschaft, German Ministry for Education and Research, Netherlands Organisation for Scientific Research, Weizmann Institute of Science, I-CORE, Initial Training Network Invisibles (Marie Curie Actions, PITNGA-2011-289442), Fundacao para a Ciencia e a Tecnologia, Region des Pays de la Loire, Knut and Alice Wallenberg Foundation, Kavli Foundation, and Istituto Nazionale di Fisica Nucleare. We are grateful to Laboratori Nazionali del Gran Sasso for hosting and supporting the XENON project.

\bibliography{inelastic_main} 

\end{document}